\documentclass{jpsj2}
\usepackage{dcolumn}% Align table columns on decimal point
\usepackage{bm}% bold math

\newcommand{\bx}{\mbox{\boldmath{$x$}}}
\newcommand{\cm}{{\mathcal M}}

\title{Statistical Mechanical Development of 
a Sparse Bayesian Classifier }

\author{ \textsc{Shinsuke Uda}$^{1}$
\thanks{E-mail address: uda@sp.dis.titech.ac.jp} and~ 
\textsc{Yoshiyuki Kabashima}$^{1}$
\thanks{E-mail address: kaba@dis.titech.ac.jp}}

\inst{$^{1}$
Department of Computational Intelligence and 
Systems Science, \\ 
Tokyo Institute of Technology, Yokohama 2268502, Japan}

\abst{
The demand for extracting rules from high dimensional 
real world data is increasing in various fields. 
However, the possible redundancy of such data sometimes 
makes it difficult to obtain a good generalization ability 
for novel samples. To resolve this problem, 
we provide a scheme that reduces the 
effective dimensions of data by pruning 
redundant components for bicategorical 
classification based on 
the Bayesian framework. First, the potential 
of the proposed method is confirmed 
in ideal situations using the replica method. 
Unfortunately, performing the scheme exactly
is computationally difficult. So, we next develop 
a tractable approximation algorithm, 
which turns out to offer nearly optimal performance 
in ideal cases when the system size is large. 
Finally, the efficacy of the developed classifier is  
experimentally examined for a real world problem of 
colon cancer classification, which shows
that the developed method can be practically useful. 
}

\kword{Bayes prediction,~Belief propagation,~Classification,
Disordered system,~Replica analysis}

\begin{document}
\maketitle

\section{Introduction}
In recent years, the demand for methods to extract rules 
from high dimensional data is increasing in the research fields of
machine learning and artificial intelligence, 
in particular, those concerning bioinformatics. 
One of the most elemental and important problems
of rule extraction is bicategorical classification 
based on a given data set~\cite{Vapnik98}. 
In a general scenario, the purpose of this task 
is to extract a certain relation between 
the input $ \bm{x} \in \mathcal{R}^{N}$, 
which is a high dimensional vector, and 
the binary output $y \in \{+1,-1\}$, which represents 
a categorical label, from a training data set 
$ D^{M} = \{(\bm{x}^1,y^{1}),\dots,(\bm{x}^\mu,y^{\mu}),
\dots (\bm{x}^M,y^{M})\}$ of $M(=1,2,\dots)$ examples. 

The Bayesian framework offers a useful guideline 
for this task. Let us assume that the relation can 
be represented by a probabilistic model defined by a 
conditional probability $P(y|\bm{x},\bm{w})$, 
where $\bm{w}$ stands for a set of adjustable 
parameters of the model. Under this assumption, 
it can be shown that for an input $\bm{x}_{M+1}$, 
the Bayesian classification 
\begin{align}
 & \hat{y}^{M+1} = \mathop{\rm argmax}_{y^{M+1}}
P(y^{M+1}|\bm{x}^{M+1},D^M)
 \label{eq:bayes_prediction}
\end{align}
minimizes the probability of misclassification
after the training set $D^M$ is observed~\cite{Iba}. 
Here $P(y^{M+1}|\bm{x}^{M+1},D^M)= 
\int d\bm{w} P(y^{M+1}|\bm{x}^{M+1},\bm{w}) 
P(\bm{w}|D^{M})$ is termed the predictive probability, and 
the posterior distribution $P(\bm{w}|D^M)$
is represented by the Bayes formula
\begin{align}
 P(\bm{w}|D^{M}) = \dfrac{P(\bm{w})P(D^{M}|\bm{w})}
 {\int d\bm{w} P(\bm{w})P(D^{M}|\bm{w})}
=\dfrac{P(\bm{w})\prod_{\mu=1}^M P(y^\mu|\bm{x}^\mu,\bm{w})}
{\int d\bm{w}P(\bm{w})\prod_{\mu=1}^MP(y^\mu|\bm{x}^\mu,\bm{w})}, 
 \label{eq:posterior_model}
\end{align}
using the traning data $D^M$ and a certain prior 
distribution $P(\bm{w})$. 

However, even this approach sometimes does not provide 
a satisfactory result for real world problems. 
A major cause of difficulty is the redundancy 
that exists in real world data. For example, let us consider 
a classification problem of DNA microarray data, 
which is a standard problem of bioinformatics. 
In such problems, while the size 
of available data sets is less than one hundred, each piece of data is typically composed 
of several thousand components, the causality or relation 
amongst which is not known in advance
~\cite{YiLi2002}. 
Simple methods that handle all the components 
usually overfit the training data, which results in 
quite a low classification performance for novel samples 
even when the Bayesian scheme of eq. (\ref{eq:bayes_prediction}) 
is performed. Therefore, when dealing with real world data, not only is the classification 
scheme itself very important but it is also important to reduce the effective dimensions,
assessing the relevance of each component. 

The purpose of this paper is to develop a scheme 
to improve the performance of the Bayesian classifier,
introducing a mechanism for eliminating
irrelevant components. The idea of the method is simple: 
in order to assess the relevance of each component of 
data, we introduce a discrete {\em pruning} parameter
$c_l \in \{0,1\}$ for each component $x_l$, 
and classify $(c_l x_l)$ instead of $\bm{x}=(x_l)$ itself. 
Components for which $c_l=0$ are 
assigned are ignored in the classification. 
Assuming an appropriate prior for $\bm{c}=(c_l)$ 
that controls the number of ignored components, 
we can introduce a mechanism to reduce 
the effective dimensions in the Bayesian 
classification (eq. (\ref{eq:bayes_prediction})), 
which is expected to lead to 
an improvement of the classification performance. 

In the literature, such 
pruning parameters have already been proposed in the research of 
perceptron learning~\cite{KuhMull94_diluted}
and linear regression problems~\cite{Iba_NMVS}. 
However, as far as the authors know, 
the potential of this method
has not been fully examined, nor have practically tractable 
algorithms been proposed for evaluating the Bayesian 
classification of eq. (\ref{eq:bayes_prediction}), 
which is computationally difficult in general. We will show 
that our scheme offers optimal 
performance in ideal situations and we provide a tractable 
algorithm that achieves a nearly optimal performance 
in such cases. For simplicity, we will here 
focus on a classifier of linear separation 
type, however, the developed 
method can be 
extended to non-linear classifiers, 
such as those based on the kernel method~\cite{BernAlex_LWK}. 

This paper is organized as follows.
In the next section, section 2, we present details of the classifier 
we are focusing on. In section 3, the performance 
of the classifier is evaluated by the replica method to 
clarify the potential of the proposed strategy. 
We show that the scheme minimizes the probability of misclassification 
for novel data
when the data includes redundant components in a certain manner. 
However, performing the scheme exactly is computationally 
difficult. In order to 
resolve this difficulty, we develop a tractable algorithm 
in section 4. We show analytically that 
a nearly optimal performance,
predicted by replica analysis, is obtained by 
the algorithm in ideal cases. 
In section 5, the efficacy for a real 
world problem of colon cancer classification is examined, 
demonstrating that the developed
scheme is competitive practically. 
The final section, section 6, is devoted to a summary.

\section{Sparse Bayesian Classifier}
The classifier that we will focus on 
is provided by a conditional probability of perceptron type
~\cite{perceptron}
\begin{equation}
 P(y|\bm{x},\bm{w},\bm{c}) = 
f\left (\frac{y}{\sqrt{N}} \sum_{l=1}^{N} c_{l} w_{l}
x_{l}  \right), 
\label{eq:model}
\end{equation}
where $c_l \in \{0,1\}$ is the pruning parameter
and the activation function $f(u)$ satisfies 
$f(u) \ge 0$ and $f(u)+f(-u)=1$ for $\forall{u} \in \mathcal{R}$. 
To introduce the pruning effect, we also use 
\begin{align}
P(\bm{w},\bm{c})={\cal V}^{-1}{ 
e^{-\sum_{l=1}^N(1-c_l)\frac{w_l^2}{2}}
\delta \left (\sum_{l=1}^N c_l -NC \right )
\delta \left (\sum_{l=1}^N c_l w_l^2 -NC \right )},
\label{eq:prior}
\end{align}
as the prior probability of 
the parameters $\bm{w}$ and $\bm{c}$, where $0\le C \le 1$ is 
a hyper parameter that controls the ratio of 
the effective dimensions. 
The factor $e^{-\sum_{l=1}^N(1-c_l)\frac{w_l^2}{2}} $ 
is included to make the normalization constant
${\cal V}=\int d\bm{w} \sum_{\bm{c}}
e^{-\sum_{l=1}^N(1-c_l)\frac{w_l^2}{2}}
\delta \left (\sum_{l=1}^N c_l -NC \right )
\delta \left (\sum_{l=1}^N c_l w_l^2 -NC \right )$ finite. 
As this {\em microcanonical} prior enforces the probability 
to vanish unless the pruning parameters $c_i \in \{0,1\}$ 
satisfy the constraint $\sum_{l=1}^N c_l =NC$, 
each specific parameter choice ignores 
certain $N(1-C)$ components of $\bm{x}$ for classification. 
These yield the posterior distribution 
$P(\bm{w},\bm{c}|D^{M})$ via the Bayes formula
\begin{align}
 & P(\bm{w},\bm{c}|D^{M}) = \dfrac{
P(\bm{w},\bm{c}) \prod_{\mu=1}^M P(y^\mu|\bm{x}^\mu,\bm{w},\bm{c})}
 {\int d\bm{w} \sum_{\bm{c}} 
P(\bm{w},\bm{c})
\prod_{\mu=1}^M P(y^\mu|\bm{x}^\mu,\bm{w},\bm{c})}, 
 \label{eq:posterior_aw}
\end{align}
which defines the Bayesian classifier as 
\begin{align}
 & \hat{y}^{M+1} = \mathop{\rm argmax}_{y^{M+1}}
 \int d\bm{w} \sum_{\bm{c}} f\left (\dfrac{y^{M+1}}{\sqrt{N}} 
 \sum_{l=1}^{N} c_{l} w_{l}x^{M+1}_l \right )
 P(\bm{w},\bm{c}|D^{M}). 
 \label{eq:estimate_y}
\end{align}
The pruning vector $\bm{c}$ eliminates irrelevant 
dimensions of data, which makes $\bm{x}$ sparse. 
Therefore, we term the classification 
scheme represented in eq. (\ref{eq:estimate_y}) 
the {\em sparse Bayesian classifier} (SBC). 

\section{Replica analysis}
To evaluate the ability of the SBC, let us assume 
the following teacher-student scenario~\cite{perceptron}. 
In this scenario, a ``teacher'' classifier is selected from 
a certain distribution $P_{\rm t}(\bm{w}_{o})$. 
For each of $M$ inputs $\bm{x}^1, \bm{x}^2,\ldots,\bm{x}^M$
which are independently generated from an 
identical distribution $P_{\rm in}(\bm{x})$, the teacher 
provides a classification label $y=\pm 1$ following 
the conditional probability 
$P(y|\bm{x},\bm{w}_{o})=f\left (\frac{y}{\sqrt{N}}\sum_{l=1}^N
w_{ol} x_l \right )$, which constitutes the training 
data set $D^M$. Then, the performance of 
the SBC, which plays the role of ``student'' in this scenario, can be 
measured by the generalization error,
which is defined as the probability of misclassification 
for a test input generated from $P_{\rm in}(\bm{x})$. 

To represent a situation 
where certain dimensions are not relevant 
for the classification label, 
let us assume that the teacher distribution is 
provided as 
\begin{align}
P_{\rm t}(\bm{w}_{o})=\prod_{l=1}^N
\left [(1-C_{\rm t}) \delta(w_{ol})+C_{\rm t} \frac{e^{-\frac{w_{ol}^2}{2}}}
{\sqrt{2 \pi}}
\right ], 
\label{teacher_dist}
\end{align}
where $C_{\rm t}$ is the ratio of the relevant dimensions 
that the student does not know. 
For simplicity, we further assume that 
the inputs are generated from a spherical distribution 
$P_{\rm in}(\bm{x})=P_{\rm sph}(\bm{x})\propto \delta \left (
|\bm{x}|^2-N \right )$, which guarantees 
that 
the correlation of $c_l w_l$ between different 
components is sufficiently weak
when the parameters $\bm{w}$ and $\bm{c}$ are
generated from the posterior $P(\bm{w},\bm{c}|D^M)$. 
As a hard constraint, $\sum_{l=1}^N c_l w_l^2
=\sum_{l=1}^N (c_lw_l)^2=NC$ is 
introduced by the microcanonical prior of eq. (\ref{eq:prior}). 
This implies that one can approximate
the {\em stability} $\Delta= \frac{y}{\sqrt{N}} 
\sum_{l=1}^N c_l w_l x_l$ as
\begin{align}
\Delta\simeq 
\sqrt{C-Q} u+\left \langle \Delta \right \rangle , 
\label{gauss_app}
\end{align}
using a Gaussian random variable $u \sim {\cal N}(0,1)$, 
where $\left \langle \cdots \right \rangle =
\int d \bm{w}\sum_{\bm{c}}P(\bm{w},\bm{c}|D^M) 
(\cdots )$ 
and $Q=\frac{1}{N}\sum_{l=1}^N \left \langle c_l w_l 
\right \rangle^2$. This 
makes it possible to evaluate eq. 
(\ref{eq:estimate_y}) as
\begin{align}
\hat{y}^{M+1}
&= {\rm sign}\left (
\int Du f \left (
\sqrt{C-Q}u +\left \langle \Delta \right \rangle 
\right )  -\frac{1}{2} \right ), 
\label{gauss_bayes}
\end{align}
where ${\rm sign}(x)=\frac{x}{|x|}$ 
for
$x \neq 0$,
and $Du= \frac{du}{\sqrt{2\pi}}e^{-\frac{u^2}{2}}$. 
Further, one can also deal with the average stability 
$\left \langle \Delta \right \rangle $ and the teacher's stability 
$\Delta^o= \frac{y}{\sqrt{N}}\sum_{l=1}^N
w_{ol}x_l$ as Gaussian random variables, the 
variances and covariance of which are given as 
\begin{align}
\overline{(\Delta^o)^2}=\frac{\bm{w}_o \cdot \bm{w}_o}{N}
\simeq C_{\rm t}, 
\quad \overline{\Delta^o\left \langle \Delta \right \rangle}
=R, \quad \overline{\left \langle \Delta \right \rangle^2}=Q, 
\label{covariance}
\end{align}
% uda revise
using $\overline{x_i x_j} = \delta_{ij}$
where $\overline{\cdots}=\int d\bm{x} P_{\rm sph}(\bm{x}) 
(\cdots )$ and $R=\frac{1}{N} \sum_{l=1}^N w_{ol} \left \langle 
c_l w_l \right \rangle$. 
This, in conjunction with the symmetry of
$y=\pm 1$ in the current system, 
indicates that the generalization error of the SBC 
can be evaluated as
\begin{align}
\epsilon^{\rm SBC}_g
=2 \int Dz
\left (1-\int Dv f
\left 
(\sqrt{C_{\rm t}-\frac{R^2}{Q}}v+\frac{R}{\sqrt{Q}}z
\right ) \right )
\Theta 
\left (
\int Du f \left (
\sqrt{C-Q}u +\sqrt{Q}z \right )  -\frac{1}{2} \right ), 
\label{Bayes_ge}
\end{align}
where $\Theta(x)=1$ for $x>0$ and $0$ otherwise, 
using the macroscopic variables $R$ and $Q$, which 
can be assessed by the replica 
method~\cite{Nishimori_InfoSP,Spinglass_beyond}.

To assess $R$ and $Q$, we evaluate the average of the
$n(=1,2,\ldots)$-th power of the partition function 
\begin{align}
 Z(D^{M}) & = \int d\bm{w}\sum_{\bm{c}}
P(\bm{w},\bm{c})\prod_{\mu=1}^{M} P(y^\mu|\bm{w},\bm{c},\bm{x}^\mu)
 \notag \\
 & = \int d\bm{w}\sum_{\bm{c}}
 \prod_{\mu=1}^M 
f\left (
\frac{y^{\mu}}{\sqrt{N}} \sum_{l}  c_{l} w_{l} x^\mu_l
\right )
{\cal V}^{-1}
e^{ - \sum_{l=1}^N \frac{1}{2} (1-c_{l})w_{l}^{2} }
 \delta
\left (\sum_{l=1}^N
 c_{l} w_{l}^{2} - CN \right ) \delta
\left (\sum_{l=1}^N c_{l} - CN \right )
\label{eq:model_Z}
\end{align}
with respect to the training data set $D^M$. 
The analytical continuation from $n=1,2,\ldots$ to 
$n \in \mathcal{R}$ under the replica symmetric (RS) ansatz 
provides the expression for the RS free energy: 
\begin{align}
&\frac{1}{N}{\left [ \ln Z(D^M) \right ]_{D^M,\bm{w}_o}}
=\frac{1}{N} \lim_{n \to 0} \frac{\partial \ln \left [ Z^n(D^M) 
\right ]_{D^M,\bm{w}_o}}{\partial n} \cr
&= \mathop{\rm Ext}_{R,Q,\hat{R},\hat{Q},F,\lambda}
\left [
2 \alpha \int Dz \int Dv
f\left (\sqrt{C_{\rm t}-\frac{R^2}{Q}}v+\frac{R}{\sqrt{Q}}z 
\right )
\ln \left (\int Du 
f\left (\sqrt{C-Q}u+\sqrt{Q}z \right )  \right ) \right . \cr
&-\hat{R}R+\frac{1}{2}{\hat{Q}Q}+\frac{1}{2}FC-\lambda C 
\cr
&+
\left . \left \langle 
\int Dz \ln 
\left [
1+\frac{1}{\sqrt{F+\hat{Q}}}
\exp \left [
\lambda + \frac{\left (\sqrt{\hat{Q}}z+\hat{R}w_o 
\right )^2 }{2(F+\hat{Q})}
\right ]
\right ]
\right \rangle_{w_o} \right ]
-\frac{1}{N}\ln {\cal V}, 
\label{RSfree}
\end{align}
where $\left [ \cdots \right ]_{D^M,\bm{w}_o}$ denotes
the average over the training set $D^M$
and the teacher distribution of eq. (\ref{teacher_dist}), 
$\mathop{\rm Ext}_{x}(\cdots)$ denotes 
extremization of $\cdots$ with respect to $x$, 
which determines $R$ and $Q$, and 
$\alpha =M/N$ and $\left \langle \cdots 
\right \rangle_{w_o}=
\int dw_o \left ((1-C_{\rm t})\delta(w_o)+C_{\rm t} \frac{e^{-\frac{w_o^2}{2}}}
{\sqrt{2 \pi}} \right ) (\cdots)$. 

\begin{figure}
\begin{minipage}{0.45\linewidth}
 \begin{center}
  \includegraphics[width=6cm,height=6cm,clip]{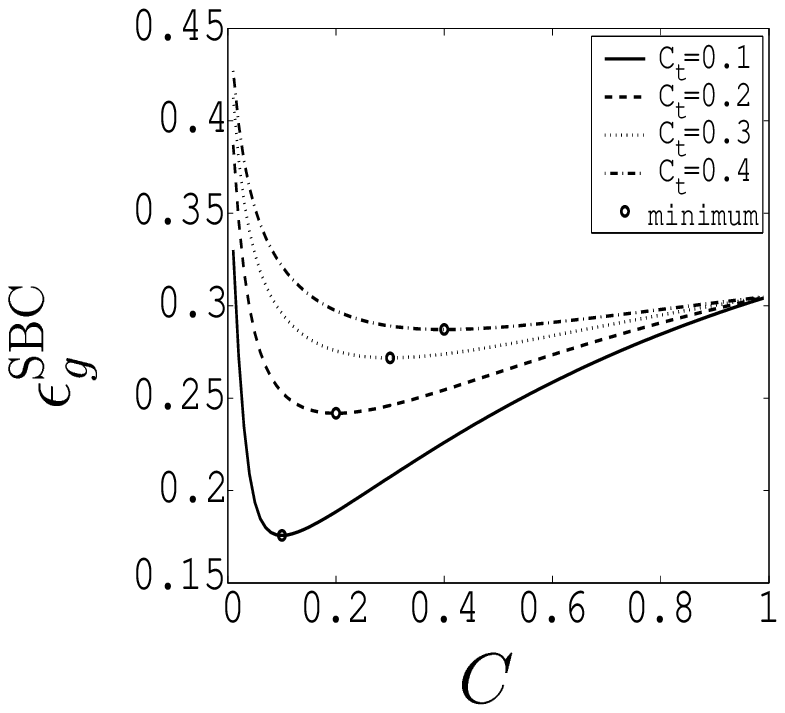}
 \end{center}
\end{minipage}
\hspace{0.5cm}
\begin{minipage}{0.45\linewidth}
 \begin{center}
  \includegraphics[width=6cm,height=6cm,clip]{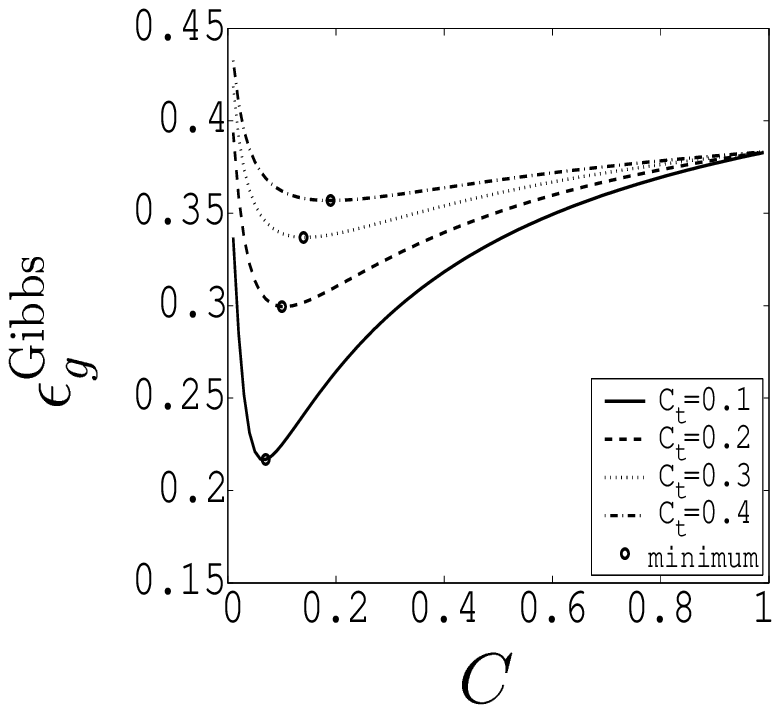}
 \end{center}
\end{minipage}
 \caption{Generalization error versus $C$ for (a) the SBC 
and (b) Gibbs learning ($\alpha=1,\kappa=0.05$).}
 %\vspace*{2mm}
 \label{fig:RS_04Nov}
\end{figure}

The generalization error $\epsilon^{\rm SBC}_g$ 
can be evaluated from eq. (\ref{Bayes_ge}) 
using $R$ and $Q$ obtained 
via the extremization problem of eq. (\ref{RSfree}), 
which is plotted in fig. \ref{fig:RS_04Nov} (a) as
a function of the hyper parameter $C$ in the case of 
\begin{align}
f(u)=f(u;\kappa)= \kappa+(1-2 \kappa) \Theta(u), 
\label{kappc_function}
\end{align}
where $\kappa ( < 1/2)$ is a non-negative constant. 
This figure shows that the developed scheme improves 
the classification ability for the assumed teacher model
of eq. (\ref{teacher_dist}) when the hyper parameter $C$ 
is appropriately adjusted. Actually, $\epsilon_g^{\rm SBC}$ 
is minimized when the hyper parameter is set to the teacher's 
value, $C=C_{\rm t}$, independent of the specific choice 
of the activation function $f(u)$. 
This is because the microcanonical prior of eq. (\ref{eq:prior})
practically coincides with the teacher model of eq. (\ref{teacher_dist}) 
when $C=C_{\rm t}$ and, therefore, the predictive 
probability $P(y^{M+1}|\bm{x}^{M+1},D^M)$ can be 
correctly evaluated in such cases. 
That the classification based on the correct predictive 
probability provides the best performance among all the 
possible strategies~\cite{Iba}, implies that 
the proposed scheme is optimal for the assumed 
teacher model if the hyper parameter is correctly tuned. 

Two things are worth discussing further. 
Firstly, the simplest replica symmetry was assumed in the above 
analysis, the validity of which should be examined. 
In fact, the RS ansatz can be broken for sufficiently small $C$, 
for which certain replica symmetry breaking (RSB) 
analysis~\cite{Spinglass_beyond} is required. 
However, at the optimal choice of hyper parameter 
$C=C_{\rm t}$, the analysis can be considered to be correct because 
this choice of prior corresponds to 
the Nishimori condition known in spin-glass research 
at which no RSB is expected to 
occur~\cite{Nishimori_line}. 
Therefore, we do not perform RSB analysis in this paper. 
Secondly, the above analysis implies that 
minimization of the generalization error 
can be used to estimate hyper parameters 
for the SBC, which will be employed 
for a real world problem in a later section. 
However, this is not necessarily the case for 
other learning strategies. 
For example, Gibbs learning, which was extensively 
examined in the last few 
decades~\cite{perceptron,OpperHaussler} 
may be used for the classification. 
In this, the classification label 
for novel data is
$\hat{y}^{\rm Gibbs}=
\mathop{\rm argmax}_{y=\pm 1}
\left \{f\left (\frac{y}{\sqrt{N}}\sum_{l=1}^N
c_l w_l x_l \right ) \right \}$,
using a pair of parameters $\bm{w}$ and $\bm{c}$
that are randomly selected from the posterior distribution of eq. 
(\ref{eq:posterior_aw}). 
The generalization error of this approach 
can be assessed as 
\begin{align}
\epsilon_g^{\rm Gibbs}=2
\int Dz \left (1-\int Dv f
\left 
(\sqrt{C_{\rm t}-\frac{R^2}{Q}}v+\frac{R}{\sqrt{Q}}z
\right ) \right )
\int Du 
\Theta \left (f \left (
\sqrt{C-Q}u +\sqrt{Q}z \right ) -\frac{1}{2} \right ), 
\label{Gibbs_ge}
\end{align}
using $R$ and $Q$. 
Although such a strategy may seem somewhat similar to 
using the SBC, the generalization error of this is not
minimized at $C=C_{\rm t}$, 
as shown in fig. \ref{fig:RS_04Nov} (b), 
which indicates that minimizing 
$\epsilon_g^{\rm Gibbs}$ is not useful for 
identifying $C$. 
This may be a reason why the determination of hyper parameters 
was not discussed in preceding work of ref. \cite{KuhMull94_diluted}, 
in which only variants of Gibbs learning were examined.

\section{Development of a BP-Based Tractable Algorithm}
The analysis in the preceding section 
indicates that the SBC of eq. (\ref{eq:estimate_y}) 
can provide optimal 
performance for a relation that is typically generated from 
eq. (\ref{teacher_dist}). Unfortunately, exact performance of 
the SBC is computationally difficult because a high-dimensional 
integration or summation with respect to $\bm{w}$ and $\bm{c}$
is required. Regarding this difficulty, recent 
studies~\cite{Mackay,Kaba_BP_SG,Kaba_cdma_belief} 
have shown that an algorithm termed {\em belief propagation}
(BP), which was developed in the information 
sciences~\cite{Gallager,Pearl_BP}, can serve as an excellent 
approximation algorithm. Therefore, let us try to construct
a practically tractable algorithm for the SBC based on BP. 

\subsection{Belief Propagation}
\begin{figure}
 \begin{center}
  \includegraphics[width=10cm,clip]{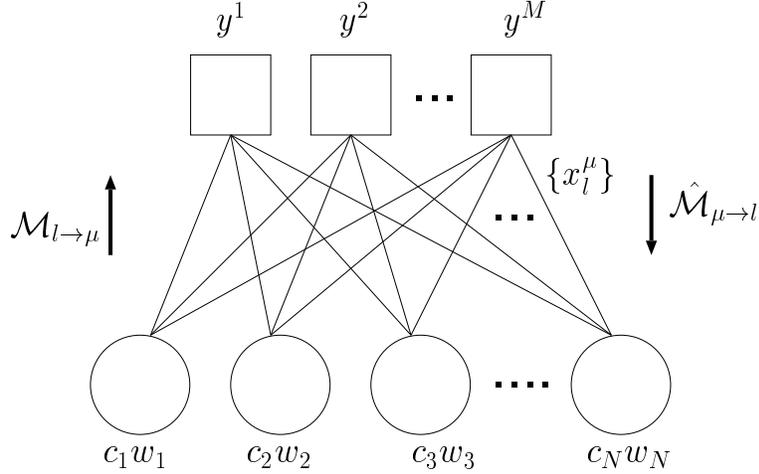}
 \end{center}
 \caption{Complete bipartite graph representing the
posterior distribution of eq. (\ref{eq:posterior_aw}).}
 \label{fig:2graph}
\end{figure}

For this, we first pictorially represent the posterior 
distribution of eq. (\ref{eq:posterior_aw}) by a complete 
bipartite graph, shown in fig. \ref{fig:2graph}.
In this figure, two types of nodes stand for 
the pairs of parameters $(w_l,c_l)$ (circle) and 
labels $y^\mu$ (square), while the edges 
connecting these nodes denote the components of the data $x_l^\mu$. 
We approximate the microcanonical prior
of eq. (\ref{eq:prior}) by 
a factorizable {\em canonical} prior as 
\begin{align}
P(\bm{w},\bm{c})
\propto \prod_{l=1}^N \exp
\left (-\frac{1}{2}(1-c_l+Gc_l)w_l^2+\lambda c_l\right ), 
\label{canonical}
\end{align}
where $G>0$ and $\lambda$ are adjustable hyper parameters. 
Then, BP is defined as an algorithm that updates 
the two types of function of $(w_l,c_l)$, which are termed 
{\em messages}, as
\begin{align}
\hat{\cm}_{\mu \to l}^{t+1}(w_l,c_l)&=\frac{
\int \prod_{j \ne l} dw_j \sum_{\bm{c}\backslash c_l}
f\left (\Delta_\mu \right )
\prod_{j \ne l } \cm_{j \to \mu}^t(w_j,c_j) 
}
{
\int d\bm{w} 
\sum_{\bm{c}}
f\left (\Delta_\mu \right )
\prod_{j \ne l } \cm_{j \to \mu}^t(w_j,c_j) 
}, 
\label{HS} \\
\cm_{l \to \mu}^t(w_l,c_l)&= 
\frac{ e^{-\frac{1}{2}(1-c_l+Gc_l)w_l^2+\lambda c_l}
\prod_{\nu \ne \mu}
\hat{\cm}_{\nu \to l}^t(w_l,c_l)}{
\int dw_l \sum_{c_l}
e^{-\frac{1}{2}(1-c_l+Gc_l)w_l^2+\lambda c_l}
\prod_{\nu \ne \mu}
\hat{\cm}_{\nu \to l}^t(w_l,c_l)},
\label{VS}
\end{align}
between the two types of nodes, where 
$\Delta_\mu=\frac{y^\mu}{\sqrt{N}}\sum_{l=1}^N c_l w_l x_l^\mu$ 
and $t$ denotes the number of updates. 
$\cm^t_{l \to \mu}(w_l,c_l)$ and 
$\hat{\cm}_{\mu \to l}^t(w_l,c_l)$ are 
approximations at the $t$-th update of 
the marginal probability of a cavity 
system in which a single element of data $(\bm{x}^\mu,y^\mu)$ 
is left out from $D^M$ and the effective influence on 
$(w_l,c_l)$ when $\bm{x}^\mu$ is newly introduced to 
the cavity system, respectively.
These provide an approximation of the posterior 
marginal at each update, 
which is termed the {\em belief}, given by
\begin{align}
P(w_l,c_l|D^M)
=\int \prod_{j \ne l}dw_j 
\sum_{\bm{c}\backslash c_l}
P(\bm{w},\bm{c}|D^M)
\simeq 
\frac{ e^{-\frac{1}{2}(1-c_l+G^tc_l)w_l^2+\lambda^t c_l}
\prod_{\mu=1}^M
\hat{\cm}_{\mu \to l}^t(w_l,c_l)}{
\int dw_l \sum_{c_l}
e^{-\frac{1}{2}(1-c_l+G^tc_l)w_l^2+\lambda^t c_l}
\prod_{\mu=1}^M
\hat{\cm}_{\mu \to l}^t(w_l,c_l)}. 
\label{belief}
\end{align}
At each update, the hyper parameters $G^t$ and $\lambda^t$ 
are determined so that the pruning 
constraints $\sum_{l=1}^N c_l =NC$ and $\sum_{l=1}^N c_lw_l^2 =NC$ 
are satisfied {\em on average} with respect to eq. (\ref{belief}), 
% uda revised
%% kaba further revised
which is valid for large $N$ due to the law of 
large numbers. 

\subsection{Gaussian Approximation and Self-Averaging Properties}
Exactly evaluating eq. (\ref{HS}) is, unfortunately, still 
difficult. In order to resolve this difficulty, 
we introduce the Gaussian approximation: 
\begin{align}
\Delta_\mu \simeq \frac{y^\mu x_l^\mu }{\sqrt{N}}c_l w_l 
+\frac{1}{\sqrt{N}}\sum_{j \ne l}{y^\mu x_j^\mu}m_{j \to \mu}^t 
+\sqrt{V_{\mu \backslash l}^t} u, 
\label{gaussian}
\end{align}
where $u \sim {\cal N}(0,1)$, 
$m_{j \to \mu}^t = \int dw_j \sum_{c_j}
(c_j w_j)\cm^t_{j \to \mu}(w_j,c_j)$
and $V_{\mu\backslash l}^t$ 
represents the variance of $\Delta_{\mu
\backslash l}= \frac{1}{\sqrt{N}}
\sum_{j \ne l}y^\mu x_j^\mu c_lw_l$. 
This approximation is likely to typically hold 
for large $N$ due to the central limit theorem when 
the parameters $(w_{j \ne l},c_{j \ne l})$ 
are generated from the cavity distribution 
$\prod_{j \ne l } \cm_{j \to \mu}^t(w_j,c_j)$
in the case when the training data $\bm{x}^\mu$ are 
independently generated from $P_{\rm sph}(\bm{x})$. 
Further, we assume that the self-averaging property 
holds for $V_{\mu\backslash l}^t$, which implies that 
$V_{\mu\backslash l}^t$ typically converges 
to its sample average independently of $\bm{x}^\mu$ and
can be evaluated using the $t$-th belief of eq. (\ref{belief}) as 
\begin{align}
V_{\mu\backslash l}^t &=
\left \langle 
\left (\Delta_{\mu
\backslash l} -\left \langle \Delta_{\mu 
\backslash l} \right \rangle_\mu^t 
\right )^2 \right \rangle_\mu^t 
= \frac{1}{N}\sum_{j,k \ne l}
x_j^\mu x_k^\mu 
\left \langle (c_jw_j-m_{j \to \mu}^t)
(c_kw_k-m_{k \to \mu}^t) \right \rangle_\mu^t \cr
&\to 
\frac{1}{N}\sum_{j,k \ne l}
\overline{x_j^\mu x_k^\mu }
\left \langle (c_jw_j-m_{j \to \mu}^t)
(c_kw_k-m_{k \to \mu}^t) \right \rangle_\mu^t 
=\frac{1}{N}\sum_{j,k \ne l} \delta_{jk}
\left \langle (c_jw_j-m_{j \to \mu}^t)
(c_kw_k-m_{k \to \mu}^t) \right \rangle_\mu^t  \cr
&=\frac{1}{N}\sum_{j \ne l}
\left (\left \langle (c_j w_j)^2 \right \rangle_\mu^t
-(m_{j \to \mu}^t)^2 \right )
\simeq \frac{1}{N} \sum_{l=1}^N 
\left (\left \langle (c_l w_l)^2 \right \rangle^t
-(m_{l }^t)^2 \right )=C-Q^t, 
\label{sa_1}
\end{align}
where $\left \langle \cdots \right \rangle_\mu^t$
and $\left \langle \cdots \right \rangle^t$
denote averages over the cavity distribution 
$\prod_{j \ne l } \cm_{j \to \mu}^t(w_j,c_j)$
and the belief of eq. (\ref{belief}), respectively, 
$m_l^t = \left \langle c_l w_l \right \rangle^t$ 
and $Q^t=\frac{1}{N}\sum_{l=1}^N (m_l^t)^2$. 
%%%%
We note that this property was once assumed to hold in 
equilibrium for similar 
systems~\cite{Mezard_Perceptron,OpperWinther97_FNN}. 
We here further assume that this 
can be extended even to the transient 
stage in BP dynamics~\cite{Kaba_BP_SG,Kaba_cdma_belief}. 
%%%%
This provides an expression of eq. (\ref{HS}):
\begin{align}
\hat{\cm}_{\mu \to l}^{t+1}(w_l,c_l)
\propto 
\exp \left [\frac{y^\mu x_l^\mu }{\sqrt{N}}a_{\mu \to l}^{t+1} 
(c_l w_l) +\frac{(y^\mu x_l^\mu)^2}{2 N}b_{\mu 
\to l}^{t+1} (c_l w_l)^2 +O(N^{-3/2})\right ],
\label{hat_to_a}
\end{align}
where 
\begin{align}
a_{\mu \to l}^{t+1}=
\frac{\partial \ln \left [\int Du f
\left (
\sqrt{C-Q^t} u+\left \langle \Delta_{\mu \backslash l}
\right \rangle_\mu^t \right )\right ]}{
\partial \left \langle \Delta_{\mu \backslash l}
\right \rangle_\mu^t }
=\frac{\int Du f^{\prime }
\left (
\sqrt{C-Q^t} u+\left \langle \Delta_{\mu \backslash l}
\right \rangle_\mu^t \right )}{
\int Du f \left (
\sqrt{C-Q^t} u+\left \langle \Delta_{\mu \backslash l}
\right \rangle_\mu^t \right )},
\label{a_mu_l}
\end{align}
\begin{align}
b_{\mu \to l}^{t+1}
& =\left ( \frac{\partial}
{\partial \left \langle \Delta_{\mu \backslash l}
\right \rangle_\mu^t } \right )^2  
\ln  \left [\int Du f
\left (
\sqrt{C-Q^t} u+\left \langle \Delta_{\mu \backslash l}
\right \rangle_\mu^t \right )\right ] \notag \\
& =\frac{\int Du f^{\prime \prime}
\left (
\sqrt{C-Q^t} u+\left \langle \Delta_{\mu \backslash l}
\right \rangle_\mu^t \right )}{
\int Du f \left (
\sqrt{C-Q^t} u+\left \langle \Delta_{\mu \backslash l}
\right \rangle_\mu^t \right )}-(a_{\mu \to l}^{t+1})^2,
\label{b_mu_l}
\end{align}
Inserting these into eq. (\ref{VS}) offers
the cavity average $m_{l \to \mu}^t$ as
\begin{align}
m_{l \to \mu}^t=\frac{
\frac{1}{\sqrt{F^t+\hat{Q}^t}}
\exp \left [
\lambda^t +\frac{(h_{l\backslash \mu}^t)^2}{2(F^t+\hat{Q}^t)}
\right ]}{
1+\frac{1}{\sqrt{F^t+\hat{Q}^t}}\exp \left [
\lambda^t +\frac{(h_{l\backslash \mu}^t)^2}{2(F^t+\hat{Q}^t)}
\right ]}
\frac{h_{l \backslash \mu}^t}{F^t+\hat{Q}^t},
\label{m_l_mu}
\end{align}
where $h_{l \backslash \mu}^t =
\frac{1}{\sqrt{N}}\sum_{\nu \ne \mu}
y^\nu x_l^\nu a_{\nu \to l}^t$ and we 
have introduced the novel macroscopic parameters 
\begin{align}
F^t&=G^t-\frac{1}{N}\sum_{\nu\ne \mu}^N
\frac{\int Du f^{\prime \prime}
\left (
\sqrt{C-Q^{t-1}} u+\left \langle \Delta_{\nu \backslash l}
\right \rangle_\nu^{t-1} \right )}{
\int Du f \left (
\sqrt{C-Q^{t-1}} u+\left \langle \Delta_{\nu \backslash l} 
\right \rangle_\nu^{t-1} \right )} \notag \\
&\simeq G^t-\frac{1}{N}\sum_{\mu=1}^N\frac{\int Du f^{\prime \prime}
\left (
\sqrt{C-Q^{t-1}} u+\left \langle \Delta_{\mu}
\right \rangle_\mu^{t-1} \right )}{
\int Du f \left (
\sqrt{C-Q^{t-1}} u+\left \langle \Delta_{\mu}
\right \rangle_\mu^{t-1} \right )},\\
\hat{Q}^t&=\frac{1}{N}
\sum_{\nu \ne \mu} (a_{\nu \to l}^{t})^2
\simeq 
\frac{1}{N}
\sum_{\mu=1}^M 
\left (\frac{\int Du f^\prime
\left (
\sqrt{C-Q^{t-1}} u+\left \langle \Delta_{\mu}
\right \rangle_\mu^{t-1} \right )}{
\int Du f \left (
\sqrt{C-Q^{t-1}} u+\left \langle \Delta_{\mu}
\right \rangle_\mu^{t-1} \right )} \right )^2,
\end{align}
assuming the further self-averaging property
\begin{align}
&\sum_{\nu\ne \mu } \frac{(y^\nu x_l^\nu)^2}{N}b_{\nu \to l}^t
\to 
\sum_{\mu=1 }^M \frac{\overline{(y^\mu x_l^\mu)^2}}{N}
b_{\mu \to l}^t 
\simeq \frac{1}{N}\sum_{\mu=1}^M b_{\mu \to l}^t \notag \\
&\simeq 
\frac{1}{N}\sum_{\mu=1}^M 
\left (\frac{\partial}
{\partial \left \langle \Delta_{\mu}
\right \rangle_\mu^{t-1} } \right )^2
\ln  \left [\int Du f
\left (
\sqrt{C-Q^{t-1}} u+\left \langle \Delta_{\mu }
\right \rangle_\mu^{t-1} \right )\right ]
\equiv \Gamma^t. 
\end{align}
In eq. (\ref{m_l_mu}), 
the adjustable hyper parameters $F^t$ and $\lambda^t$ are 
determined so that the average pruning conditions 
\begin{align} 
 & \frac{1}{N} \sum_{l=1}^N \langle c_{l} w_{l}^{2} \rangle^t = 
\frac{1}{N}\sum_{l=1}^N \dfrac{ \frac{1}{\sqrt{F^t+\hat{Q}^t}} 
 \exp \left[ \lambda^t + \frac{(h_{l}^t)^{2}}{2(F^t+\hat{Q}^t)} 
\right] }
 { 1+ \frac{1}{\sqrt{F^t+\hat{Q}^t}} 
\exp \left[ \lambda^t + \frac{(h_l^t)^{2}}
{2(F^t+\hat{Q}^t)} \right] }
 \left( \frac{1}{F^t+\hat{Q}^t} + 
\frac{ (h_{l}^t)^{2} } { (F^t+\hat{Q}^t)^{2} } \right)= C, 
 \label{eq:aw2_C} \\
 & \frac{1}{N} \sum_{l=1}^N \langle c_{l} \rangle^t = 
\frac{1}{N}\sum_{l=1}^N 
\dfrac{ \frac{1}{\sqrt{F^t+\hat{Q}^t}} 
 \exp \left[ \lambda^t + \frac{(h_{l}^t)^{2}}{2(F^t+\hat{Q}^t)} 
\right] }
 { 1+ \frac{1}{\sqrt{F^t+\hat{Q}^t}} \exp \left[ \lambda^t + 
 \frac{(h_{l}^t)^{2}}{2(F^t+\hat{Q}^t)} \right] } = C, 
 \label{eq:a_C}
\end{align}
hold for eq. (\ref{belief}) at each update, 
where $h_l^t= \frac{1}{\sqrt{N}}\sum_{\mu=1}^M 
y^\mu x_l^\mu a_{\mu \to l}^t$. 

Notice that eqs. (\ref{a_mu_l})-(\ref{eq:a_C}) can 
be used as a computationally tractable algorithm for 
assessing the SBC. 
For this, we evaluate the posterior average $m_l=\left \langle 
c_l w_l \right \rangle$, plugging eq. 
(\ref{hat_to_a}) into eq. (\ref{belief})
using eqs. (\ref{a_mu_l}) and (\ref{b_mu_l}), 
which provides 
\begin{align}
m_l^t=\dfrac{ \frac{1}{\sqrt{F^t+\hat{Q}^t}} 
 \exp \left[ \lambda^t + \frac{(h_{l}^t)^{2}}{2(F^t+\hat{Q}^t)} 
\right] }
 { 1+ \frac{1}{\sqrt{F^t+\hat{Q}^t}} \exp \left[ \lambda^t + 
 \frac{(h_{l}^t)^{2}}{2(F^t+\hat{Q}^t)} \right] }
\frac{h_l^t}{F^t+\hat{Q}^t}. 
\label{m_l}
\end{align}
This makes it possible to evaluate the SBC of eq. (\ref{eq:estimate_y}) using 
the Gaussian approximation as
\begin{align}
\hat{y}={\rm sign}\left (Du f(\sqrt{C-Q^t}u+\left \langle 
\Delta \right \rangle^t )-\frac{1}{2} \right ), 
\label{tractable_SBC}
\end{align}
for an element of data $\bm{x}$ that is newly generated 
from $P_{\rm sph}(\bm{x})$ at each update.

\subsection{Further Reduction of Computational Cost}
The necessary cost of performing the above procedure 
is $O(N^2M)$ per update, which can be further reduced. 
In order to save the computational cost, we represent 
$\left \langle 
\Delta_{\mu \backslash l} \right \rangle_\mu^t$ 
and $h_{l \backslash \mu}^t$ as
\begin{align}
\left \langle \Delta_{\mu \backslash l} \right \rangle_\mu^t
&\simeq
\left \langle \Delta_\mu \right \rangle^t-\frac{1}{\sqrt{N}}
\sum_{l=1}^N y^\mu x_l^\mu 
\frac{\partial m_l^t}
{\partial h_l^t} \left (\frac{y^\mu x_l^\mu}{\sqrt{N}} 
a_{\mu \to l}^t \right ) -\frac{y^\mu x_l^\mu }{\sqrt{N}}
m_{l \to \mu}^t 
\simeq\left \langle \Delta_\mu \right \rangle^t
-\Xi^t a_\mu^t -
\frac{y^\mu x_l^\mu }{\sqrt{N}}m_{l}^t, \\
h_{l \backslash \mu}^t &\simeq \theta_l^t-\frac{1}{\sqrt{N}}
\sum_{\mu=1}^M y^\mu x_l^\mu\frac{\partial a_\mu^t}
{\partial \left \langle \Delta_\mu \right \rangle_\mu^t}
\left (\frac{y^\mu x_l^\mu}{\sqrt{N}} 
m_{l \to \mu}^{t-1}\right )-\frac{y^\mu x_l^\mu}{\sqrt{N}}
a_{\mu \to l}^t
\simeq \theta_l^t -\Gamma^t m_l^{t-1} -\frac{y^\mu x_l^\mu}{\sqrt{N}}
a_{\mu}^t, 
\end{align}
using the singly-indexed variables 
$m_l^t$,
$a_\mu^t=\frac{\partial \ln \left [\int Du f(\sqrt{C-Q^{t-1}}+
\left \langle \Delta_\mu \right \rangle^{t-1}_\mu )
\right ]}
{\partial \left \langle \Delta_\mu \right \rangle^{t-1}_\mu }$, 
and
$\left \langle \Delta_\mu \right \rangle^t
=\frac{1}{\sqrt{N}}\sum_{l=1}^N
y^\mu x_l^\mu m_l^t$,
where 
$\theta_l^t=\frac{1}{\sqrt{N}}\sum_{\mu=1}^M
y^\mu x_l^\mu a_\mu^t$ and  
$\Xi^t\equiv C-Q^t$. 
Using these, the algorithm developed above can be summarized as 
\begin{align}
a_\mu^{t+1}&=\frac{\partial \ln 
\left [
\int Du f\left (\sqrt{C-Q^t} u+\left \langle \Delta_\mu 
\right \rangle^t -\Xi^ta_\mu^t \right ) \right ]}{
\partial \left \langle \Delta_\mu 
\right \rangle^t }, 
\label{low_cost_a}\\
m_l^t&=
\frac{\partial \ln \left [
1+ \frac{1}{\sqrt{F^t+\hat{Q}^t}}
\exp\left [\lambda^t+
\frac{\left (\theta_l^t-\Gamma^t m_l^{t-1} \right )^2}
{2(F^t+\hat{Q}^t)}
\right ] \right ]}{
\partial \theta_l^t}.
\label{low_cost_m}
\end{align}
This version may be useful in analyzing 
relatively higher dimensional data, 
as the computational cost per update 
is reduced from $O(N^2 M)$ to $O(NM)$.

\subsection{Performance Analysis and Link to the RS Solution}
To investigate the performance of the 
BP-based algorithm, let us describe its behavior
using macroscopic variables, such as 
$Q^t=\frac{1}{N}\sum_{l=1}^N (m_l^t)^2$ and 
$R^t=\frac{1}{N}\sum_{l=1}^N w_{ol}m_l^t$,
in the thermodynamic limit 
$N,M \to \infty$, keeping $\alpha \sim O(1)$. 
We will perform the analysis based on 
the naive expression of the algorithm in eqs. (\ref{a_mu_l})-(\ref{eq:a_C}), since eqs. (\ref{low_cost_a}) 
and (\ref{low_cost_m}) are just a cost-saving 
version of this naive algorithm and, 
therefore, their behavior is identical. 

For this purpose, we first assume that the self-averaging 
properties 
\begin{align}
\frac{1}{N}\bm{m}_\mu^t \cdot \bm{m}_\mu^t &\simeq
\frac{1}{N}\bm{m}^t \cdot \bm{m}^t =Q^t, 
\label{sa_Q}\\
\frac{1}{N}\bm{w}_o \cdot \bm{m}_\mu^t &\simeq 
\frac{1}{N}\bm{w}_o \cdot \bm{m}^t =R^t, 
\label{sa_R}
\end{align}
hold for the macroscopic variables. 
That the training data $\bm{x}^\mu$ are independently 
drawn from $P_{\rm sph}(\bm{x})$,
in conjunction with the central limit theorem, implies that 
the pair of $\left \langle \Delta_{\mu } 
\right \rangle_\mu^t$ and the teacher's stability 
$\Delta_\mu^o=\frac{y^\mu}{\sqrt{N}}\sum_{l=1}^N
x_l^\mu w_{ol}$ can be treated as zero-mean 
Gaussian random variables, the variances and covariance 
of which are
\begin{align}
\left [(\Delta^o_\mu )^2\right ]_{D^M}
=\frac{\bm{w}_o \cdot \bm{w}_o}{N}
\simeq C_{\rm t}, 
\quad \left [\Delta^o_\mu\left \langle \Delta_{\mu  
}
\right \rangle_\mu^t \right ]_{D^M}
\simeq R^t, 
\quad 
\left [
\left (\left \langle \Delta_
{\mu } \right \rangle_\mu^t \right )^2
\right ]_{D^M}
\simeq Q^t. 
\label{variance_covariance2}
\end{align}
This makes it possible to represent these variables as
\begin{align}
\Delta^o_\mu \simeq \sqrt{C_{\rm t}-\frac{(R^t)^2}{Q^t}}v +
\frac{R^t}{\sqrt{Q^t}} z, \label{teacher_delta} \\
\left \langle \Delta_{\mu }\right\rangle_\mu^t
\simeq \sqrt{C-Q^t}u +\sqrt{Q^t} z, 
\label{student_delta}
\end{align}
using three independent 
Gaussian random variables $u,v,z \sim {\cal N}(0,1)$, 
which, in conjunction with the self-averaging property, 
indicates that the macroscopic properties of the cavity field 
$h_{l \backslash \mu}^{t+1}=\frac{1}{\sqrt{N}}
\sum_{\nu \ne \mu} {y^\nu}x_l^\nu a_{\nu\to l}^{t+1}$
can be characterized 
independently of $l$ and $\mu$ as 
\begin{align}
&\frac{1}{N} \sum_{l=1}^N (h_{l \backslash \mu}^{t+1})^2 
 = \frac{1}{N^2} \sum_{l=1}^N 
\sum_{\mu,\nu}
\left [y^\mu y^\nu x_l^\mu x_l^\nu  
a_{\mu \to l}^{t+1}a_{\nu \to l}^{t+1}\right ]_{D^M} \notag \\
&\simeq \frac{1}{N^2}\sum_{l=1}^N 
\sum_{\mu=1}^M \left [ (a^{t+1}_{\mu \to l})^2 \right]_{D^M}
\simeq \frac{1}{N^2}\sum_{l=1}^N \sum_{\mu=1}^M 
\left [(a^{t+1}_{\mu})^2 \right ]_{D^M} \notag \\
&\simeq 
 2 \alpha \int Dz 
Dv f\left (\sqrt{C_{\rm t}-\frac{(R^t)^2}{Q^t}}v +
\frac{R^t}{\sqrt{Q^t}} z \right )
\left (
\frac{
\int Du f^\prime \left (
\sqrt{C-Q^t}u+\sqrt{Q^t} z \right )}
{\int Du f \left (
\sqrt{C-Q^t}u+\sqrt{Q^t} z \right )}
\right )^2\simeq \hat{Q}^{t+1}, \label{nohat_to_hatQ} \\
&\frac{1}{N}\sum_{l=1}^N w_{ol} h_{l \backslash \mu}^{t+1}
 = \frac{1}{N} \sum_{l=1}^N 
\sum_{\nu \neq \mu}^M
\left [\frac{y^\nu}{\sqrt{N}}x_l^\nu w_{ol} 
a_{\nu \to l}^{t+1} \right ]_{D^M} \notag \\
&\simeq \frac{1}{N}\sum_{\mu=1}^M 
\left [\Delta_\mu^o a_{\mu}^{t+1}  \right ]_{D^M}
-\frac{1}{N}\sum_{\mu=1}^M\sum_{l=1}^N  w_{ol}m_{l \to \mu}^t
\left [(y^\mu x_l^\mu)^2 \frac{\partial a_{\mu}^{t+1}
}{\partial \left \langle \Delta_{\mu} \right \rangle_\mu^t}
\right ]_{D^M} \notag \\ 
&\simeq 2\alpha \int Dz Dv 
\left (\sqrt{C_{\rm t}-\frac{(R^{t})^2}{Q^t}}v+\frac{R^t}{\sqrt{Q^t}}z 
\right )f\left (\sqrt{C_{\rm t}-\frac{(R^{t})^2}{Q^t}}v+
\frac{R^t}{\sqrt{Q^t}}z \right )
\frac{\int Du f^\prime \left (
\sqrt{C-Q^t}u+\sqrt{Q^t} z \right )}
{\int Du f \left (
\sqrt{C-Q^t}u+\sqrt{Q^t} z \right )} \notag \\
&-2 \alpha R^t \int Dz Dv 
f \left (\sqrt{C_{\rm t}-\frac{(R^{t})^2}{Q^t}}v+
\frac{R^t}{\sqrt{Q^t}}z \right )
\frac{\partial }{\partial (\sqrt{Q^t}z)}
\left (
\frac{\int Du f^\prime \left (\sqrt{C-Q^t}u+\sqrt{Q^t} z \right )}
{\int Du f \left (\sqrt{C-Q^t}u+\sqrt{Q^t}z \right )} \right ) \notag \\
&= 2\alpha C_{\rm t} \int Dz Dv 
f^\prime \left (\sqrt{C_{\rm t}-\frac{(R^{t})^2}{Q^t}}v+
\frac{R^t}{\sqrt{Q^t}}z \right )
\frac{
\int Du f^\prime \left (
\sqrt{C-Q^t}u+\sqrt{Q^t} z \right )}
{\int Du f \left (
\sqrt{C-Q^t}u+\sqrt{Q^t} z \right )}=C_{\rm t}
\hat{R}^{t+1}, 
\label{nohat_to_hatR}
\end{align}
where the prefactor $2$ in eqs. (\ref{nohat_to_hatQ})
and (\ref{nohat_to_hatR}) has its root in the two 
possibilities of the label $y=\pm 1$. 
On the other hand, 
these equations mean that the cavity fields, which can also be 
treated as Gaussian random variables since $x_l^\mu$ are 
zero-mean and almost uncorrelated random variables, can be 
represented as 
\begin{align}
h_{l\backslash \mu}^t=\sqrt{\hat{Q}^t}z+\hat{R}^tw_{ol}, 
\label{cavity}
\end{align}
where $z \sim {\cal N}(0,1)$, which yields
\begin{align}
Q^t &\simeq \frac{1}{N} \sum_{l=1}^N (m_{l \to \mu}^t)^2 
=\frac{1}{N}
\sum_{l=1}^N
\left (\frac{
\frac{1}{\sqrt{F^t+\hat{Q}^t}}
\exp \left [
\lambda^t +\frac{(h_{l\backslash \mu}^t)^2}{2(F^t+\hat{Q}^t)}
\right ]}{
1+\frac{1}{\sqrt{F^t+\hat{Q}^t}}\exp \left [
\lambda^t +\frac{(h_{l\backslash \mu}^t)^2}{2(F+\hat{Q}^t)}
\right ]}
\frac{h_{l \backslash \mu}^t}{F+\hat{Q}^t} \right )^2 \cr
&= \int Dz \left \langle
\left (\frac{
\frac{1}{\sqrt{F^t+\hat{Q}^t}}
\exp \left [
\lambda^t + \frac{(\sqrt{\hat{Q}^t}z+\hat{R}^tw_o)^2}
{2 (F^t+\hat{Q}^t)}
\right ]}{
1+\frac{1}{\sqrt{F^t+\hat{Q}^t}}
\exp \left [
\lambda^t + \frac{(\sqrt{\hat{Q}^t}z+\hat{R}^tw_o)^2}
{2 (F^t+\hat{Q}^t)}
\right ]} 
\frac{\sqrt{\hat{Q}^t}z+\hat{R}^tw_o}{
F^t+\hat{Q}^t}
\right )^2 \right \rangle_{w_o}, 
\label{hat_to_nohatQ} \\
R^t&\simeq \frac{1}{N}\sum_{l=1}^N w_{ol}m_{l \to \mu}^t
\simeq 
\frac{1}{N}
\sum_{l=1}^N
\frac{
\frac{1}{\sqrt{F^t+\hat{Q}^t}}
\exp \left [
\lambda^t +\frac{(h_{l\backslash \mu}^t)^2}{2(F^t+\hat{Q}^t)}
\right ]}{
1+\frac{1}{\sqrt{F^t+\hat{Q}^t}}\exp \left [
\lambda^t +\frac{(h_{l\backslash \mu}^t)^2}{2(F^t+\hat{Q}^t)}
\right ]}\frac{w_{ol}h_{l \backslash \mu}^t}{F^t+\hat{Q}^t}  \cr
&=\frac{\hat{R}^t}{F^t+\hat{Q}^t}
\int Dz \left \langle 
\frac{
\frac{1}{\sqrt{F^t+\hat{Q}^t}}
\exp \left [
\lambda^t + \frac{(\sqrt{\hat{Q}^t}z+\hat{R}^tw_o)^2}
{2 (F^t+\hat{Q}^t)}
\right ]}{
1+\frac{1}{\sqrt{F^t+\hat{Q}^t}}
\exp \left [
\lambda^t + \frac{(\sqrt{\hat{Q}^t}z+\hat{R}^tw_o)^2}
{2 (F^t+\hat{Q}^t)}
\right ]} w_o^2 \right \rangle_{w_o}, 
\label{hat_to_nohatR} 
\end{align}
where $F^t$ and $\lambda^t$ are determined so that 
the pruning conditions 
\begin{align}
&\int Dz \left \langle 
\frac{
\frac{1}{\sqrt{F^t+\hat{Q}^t}}
\exp \left [
\lambda^t + \frac{(\sqrt{\hat{Q}^t}z+\hat{R}^tw_o)^2}
{2 (F^t+\hat{Q}^t)}
\right ]}{
1+\frac{1}{\sqrt{F^t+\hat{Q}^t}}
\exp \left [
\lambda^t + \frac{(\sqrt{\hat{Q}^t}z+\hat{R}^tw_o)^2}
{2 (F^t+\hat{Q}^t)}
\right ]} 
\left (\frac{1}{F^t+\hat{Q}^t}+
\left (\frac{\sqrt{\hat{Q}^t}z+\hat{R}^t w_o}
{F^t+\hat{Q}^t} \right )^2 \right )
\right \rangle_{w_o}=C,
\label{pruning1}\\
&\int Dz \left \langle 
\frac{
\frac{1}{\sqrt{F^t+\hat{Q}^t}}
\exp \left [
\lambda^t + \frac{(\sqrt{\hat{Q}^t}z+\hat{R}^tw_o)^2}
{2 (F^t+\hat{Q}^t)}
\right ]}{
1+\frac{1}{\sqrt{F^t+\hat{Q}^t}}
\exp \left [
\lambda^t + \frac{(\sqrt{\hat{Q}^t}z+\hat{R}^tw_o)^2}
{2 (F^t+\hat{Q}^t)}
\right ]} \right \rangle_{w_o}=C,
\label{pruning2} 
\end{align}
hold for the $t$-th update.

\begin{figure}[t]
 \begin{center}
  \includegraphics[width=6cm,clip]{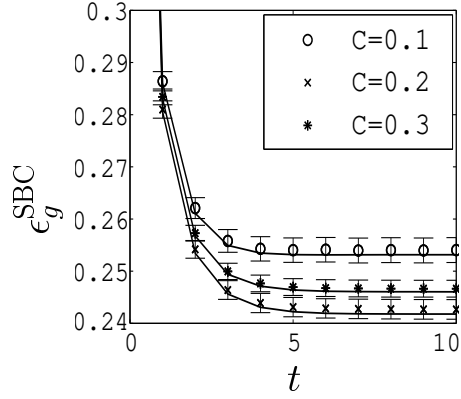}
 \end{center}
\caption{Generalization error of SBC provided
by the BP-based algorithm (\ref{low_cost_a}) 
and (\ref{low_cost_m}) after $t$th update. 
Markers were obtained from $100$ experiments for 
a transfer function $f(u;\kappa=0.05)=0.05+0.9\Theta(u)$
in the case of 
$C_{\rm t}=0.2$ 
varying $C=0.1, 0.2$ and $0.3$.
Error bars indicate 95\% confidence intervals. 
Lines represent the theoretical prediction assessed 
by eqs. (\ref{nohat_to_hatQ})-(\ref{pruning2}), which 
exhibits excellent consistency with the experimental results. 
}
 \label{time_evolution_eg}
\end{figure}

%%%%%%%%%%%%
In fig. \ref{time_evolution_eg}, experimentally obtained 
time evolution of the BP-based algorithm (\ref{low_cost_a}) 
and (\ref{low_cost_m}) is compared with 
its theoretical prediction assessed by 
eqs. (\ref{nohat_to_hatQ})-(\ref{pruning2}), 
which exhibits excellent consistency. 
This validates the macroscopic analysis provided above. 
%%%%%%%%%%%%
It is worth noting that the stationary conditions of 
eqs. (\ref{nohat_to_hatQ})-(\ref{pruning2}) are
identical to those of the saddle point of the RS free energy of eq.
(\ref{RSfree}). This implies that the BP-based algorithm 
provides a nearly optimal solution in a practical
time scale in assumed ideal situations 
of large system size if the hyper parameter 
$C_{\rm t}$ is correctly estimated. 

\begin{figure}
 \begin{center}
  \includegraphics[width=6cm,height=6cm,clip]{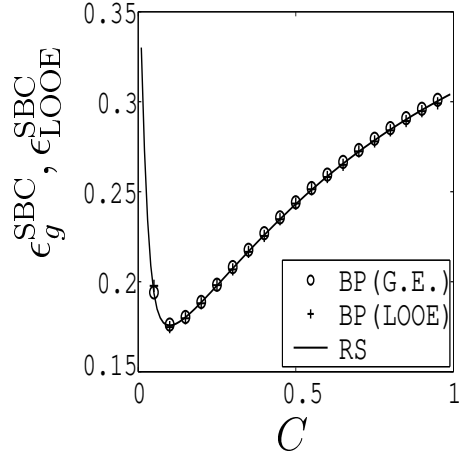}
 \end{center}
 \caption{Comparison between $\epsilon_g^{\rm SBC}$ ($\circ$)
and $\epsilon_{\rm LOOE}^{\rm SBC}$ ($+$) for 
$C_{\rm t}=0.1$,$\kappa = 0.05$ and $\alpha=1$. 
$\epsilon_g^{\rm SBC}$ is evaluated by replica analysis 
under the replica symmetric ansatz (Line),
while $\epsilon_{\rm LOOE}^{\rm SBC}$
is experimentally obtained using eqs. (\ref{low_cost_a}) 
and (\ref{low_cost_m}) for $N,M=2000$. }
 \label{fig:RS_BPart_04Nov}
\end{figure}

The replica analysis in the previous section
indicates that $C_{\rm t}$ can be estimated 
by minimizing the generalization error 
$\epsilon_g^{\rm SBC}$. 
The leave-one-out error (LOOE), which is represented as 
\begin{align}
\epsilon_{\rm LOOE}^{\rm SBC}=
\frac{1}{M}\sum_{\mu=1}^M
\Theta\left (\frac{1}{2}-
\int Du f
\left (\sqrt{C-Q}u+\left \langle 
\Delta_\mu \right \rangle_\mu \right ) \right ), 
\label{LOOE}
\end{align}
in the current case, is frequently 
used as an estimate of $\epsilon_g^{\rm SBC}$
for practical applications, since it can be evaluated 
from only the training set $D^M$. 
The algorithm is also useful for 
assessing the LOOE of eq. (\ref{LOOE}), since this 
computes all the cavity 
stabilities $\left \langle \Delta_\mu \right \rangle_\mu$ 
at each update, which saves the cost of relearning 
in assessing the LOOE~\cite{OpperWinther97_FNN}. 
Fig. \ref{fig:RS_BPart_04Nov} shows a comparison 
between $\epsilon_g^{\rm SBC}$ and 
$\epsilon_{\rm LOOE}^{\rm SBC}$ for 
an activation function, as given in eq. (\ref{kappc_function}),
in the case of $N,M=2000$. 
It shows that the LOOE (\ref{LOOE}) can be used
in practical applications for determining the necessary 
hyper parameters using only given data. 

%%%%%%%%% revision for A)  %%%%%%%%%%%%%%
The BP-based algorithm of eqs. (\ref{low_cost_a}) 
and (\ref{low_cost_m}) is developed and analyzed
under the self-averaging assumption, 
which is valid when each data $\bm{x}^\mu$ 
is independently sampled from the spherical 
distribution $P_{\textrm{sph}} (\bm{x})$.
Unfortunately, raw real world data do not necessarily 
obey such distributions, which may deteriorate
the approximation accuracy of the developed algorithm. 
A simple approach to handle this problem is 
to make statistical properties of the data set 
get closer to those of samples from $P_{\textrm{sph}} (\bm{x})$
by linearly transforming $\bm{x}^\mu$ so that 
$\frac{1}{M}\sum_{\mu=1}^M \bx^\mu=0$ and 
$\frac{1}{M}\sum_{\mu=1}^M x_i^\mu x_j^\mu 
\simeq \delta_{ij}$ hold with 
keeping $|\bm{x}^\mu|$ fixed to the square 
root of the dimensionality. 
Such an approach is often termed {\em whitening}, 
the efficacy of which will be experimentally examined 
in the next section. 
%%%%%%%%%%%%%%%%%%%%%%%%%%%%%%%%%%%%%%%%%%%%%%%%%%%

\section{Application to a Real World Problem}
To examine the practical significance of the SBC, we applied 
it to a real world problem. 
We considered the task of distinguishing cancer from normal tissue 
using microarray data 
of $N=2000$ dimensions~\cite{YiLi2002}. 
The data was sampled from $M=62$ tissues, 
20 and 42 tissues of which were classified as normal and cancerous,
respectively. 

We employed eq. (\ref{kappc_function}) as the activation function. 
The data set $D^M=\{(\bm{x}^\mu,y^\mu)\}$~\cite{Alon99} 
was pre-processed so that $\frac{1}{M} \sum_{\mu=1}^M
\bm{x}^\mu = 0$ and $|\bm{x}^\mu| = \sqrt{N}$ held. 
The data set was randomly divided into 
training and test sets, which were composed of 42 and 
20 tissues, respectively. For a given training set, 
the hyper parameters $C$ and $\kappa$ 
were determined from the possibilities of 
$C=\{2.5\times 10^{-3},5.0\times
10^{-3},1.0\times 10^{-2},
5.0\times 10^{-2},0.1,0.2,0.4,0.6,0.8\}$
and $\kappa=\{0.1,0.2,0.3,0.4\}$,
so as to minimize eq. (\ref{LOOE}) for the training set. 
After determining $C$ and $\kappa$, the generalization 
error was measured for the test set. 
We repeated this experiment 200 times, 
redividing the data set. 

The results are shown in table \ref{table:colon}.
The conventional Fisher discriminant analysis (FDA)~\cite{Fisher} 
and Sparse Bayesian learning (SBL)~\cite{Tipping01_SBL}, 
which selects a sparse model using a certain prior, 
termed the automatic relevance determination (ARD) 
prior~\cite{Mackay_Bayes_ARD,Neal_Bayes96}, 
are presented for comparison. 
%%%%%%
It is apparent that the FDA, which does not have 
a mechanism to reduce effective dimensions, 
exhibits a significantly lower generalization ability 
than the other two schemes. 
%%%%%%
%%%%%% Reply to comment B)
This is also supported by Welch's test, 
which is a standard method to examine statistical 
significance of difference of averages between two groups, 
although the standard deviation of FDA 
is smaller than those of the others. 
%%%%%%
On the other hand, although the average 
generalization error of the SBC is smaller 
than that of SBL, the standard deviations are large, which 
prevents us from clearly judging 
the superiority of the SBC. 

In order to resolve this difficulty, 
we examined how many times the SBC provided a smaller 
generalization error than SBL in the 200 experiments. 
The number of times that 
the SBC offered smaller, equal and larger errors 
than SBL were 99, 36 and 65, respectively. 
A one-sided binomial test was applied to this result under 
the null hypothesis that there is no difference 
of the generalization ability 
between SBC and SBL ignoring the tie data, 
which yields 
$\frac{|99-(200-36)\times \frac{1}{2}|}{
\sqrt{(200-36)\times \frac{1}{2} \times \frac{1}{2}}} 
= 2.65\cdots > u(0.05)=1.64\cdots$
under the normal approximation. 
%%%%%%%%
This implies that 
the difference between 
SBC and SBL is statistically significant 
with a confidence level of $95$\% and, therefore,
the SBC has a higher generalization ability. 
%%%%%%%%

%%%%%%%%%%%%%%%%%%%%%%%
\begin{table}[t]
\caption
{Classification result}
 \label{table:colon}
\begin{center}
\begin{tabular}{c|cc}
 &generalization error (\%) & standard deviation \\ 
\hline	
SBC & 23.5 & 10.2 \\
SBL & 24.3 & 10.7 \\
FDA & 39.6 & 8.39 \\
\end{tabular}
\end{center}
\end{table}

\begin{figure}
 \begin{center}
  \includegraphics[width=6cm,height=6cm,clip]{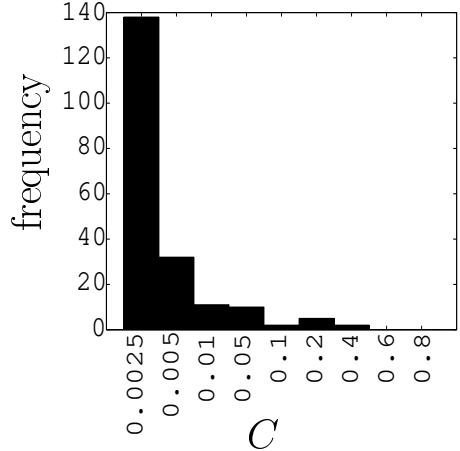}
 \end{center}
 \caption{Histogram of selected $C$}
 \label{fig:C-freq_02Dec}
\end{figure}

\begin{figure}
 \begin{center}
  \includegraphics[width=6cm,height=6cm,clip]{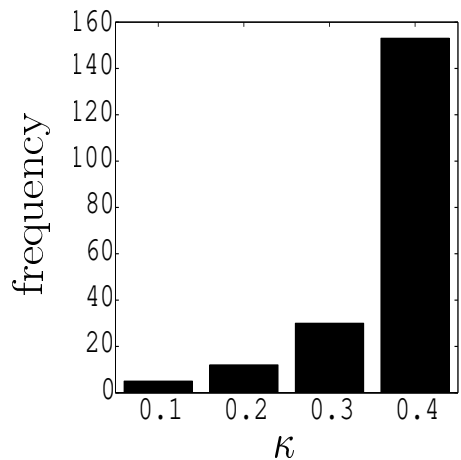}
 \end{center}
 \caption{Histogram of selected $\kappa$}
 \label{fig:k-freq_03Dec}
\end{figure}

Histograms of selected values of $C$ and $\kappa$
are shown in figs. \ref{fig:C-freq_02Dec} 
and \ref{fig:k-freq_03Dec}, respectively. 
They indicate that $\kappa$ has a statistically 
greater fluctuation. For reference, we
performed experiments fixing $\kappa$ to the 
most frequent value, $\kappa=0.4$, which reduced
the average and standard deviation of the generalization 
error of SBC to 21.0(\%) and 9.72, respectively. 
This determination by the {\em resampling} technique
is an alternative scheme for estimating hyper parameters. 
Although performing it naively requires a greater 
computational cost than minimizing the LOOE, 
a recently proposed analytical approximation 
method~\cite{OpperMarzahn} may be promising 
for reducing computational cost, 
and will be the subject of future work.

In the algorithm we have developed, we assumed 
a self-averaging property. This assumption 
may give a good match with {\em whitened} 
data, i.e., data for which the dimensionality is reduced 
from $N=2000$ to $M=62$ by a linear transformation so that 
$\frac{1}{M}\sum_{\mu=1}^M \bm{x}^\mu=0$
and $\frac{1}{M}\sum_{\mu=1}^M x_i^\mu x_j^\mu =\delta_{ij}$
hold in the reduced space. 
We also carried out the above experiments for 
the whitened data fixing $\kappa$ to $0.4$, 
finding that the average and standard deviation of 
the generalization error of the SBC were 
reduced to 16.3(\%) and 6.20, respectively. 
However, such an approach may not be preferred 
because it becomes difficult to interpret 
the implications of the result 
as the original meaning of variables is lost by 
the linear transformation. 

\section{Summary}
In summary, we have developed 
a classifier termed the sparse Bayesian classifier (SBC) 
that eliminates irrelevant components 
in high-dimensional data $\bm{x} \in \mathcal{R}^N$
by multiplying discrete variables $c_{l} \in \{0,1\},1 
\leq l \leq N$ for each dimension $l$
following the Bayesian framework. 
The efficacy of the SBC was confirmed 
by the replica method for the target rules of a certain type. 
Unfortunately, exactly evaluating the SBC is computationally 
difficult. In order to resolve this difficulty, 
we have also developed a computationally 
tractable approximation algorithm for the SBC 
based on belief propagation (BP).
It turns out that the developed BP-based algorithm 
provides a result consistent with 
that of replica analysis for ideal situations, 
which implies that a nearly optimal performance 
can be obtained in a practical time scale 
in such ideal cases. 
Finally, the significance of the SBC to real world 
applications was experimentally validated 
for a problem of colon cancer classification.

In this paper, the classifier was developed for 
minimizing the generalization error. 
Identifying relevant components 
from a given data set may be another purpose of 
the classification analysis. Designing a classifier 
for this purpose is currently under way. 

\section*{Acknowledgment}
This work was partially supported by 
Grant-in-Aid No.~14084206 from MEXT, Japan (YK). 

%\bibliographystyle{apsrev.bst}
%\bibliography{prune_p1.bib}% Produces the bibliography via BibTeX.

\end{document}